\documentclass{epl}

\usepackage{amsmath}
\usepackage{amsfonts}
\usepackage{amssymb}%

\title{Shape and scaling of moving step bunches}
\author{V. Popkov\inst{1} \and J. Krug\inst{1,2}}
\institute{
  \inst{1} Institut f\"ur Theoretische Physik, Universit\"at zu K\"oln, 
Germany \\
  \inst{2} Laboratory of Physics, Helsinki University of Technology,
Finland
}

\begin{document}

\maketitle

\begin{abstract}
We study step bunching under conditions of attachment/detachment limited
kinetics in the presence of a deposition or
sublimation flux, which leads to bunch motion. Analysis of the 
discrete step dynamics reveals that the bunch velocity is inversely
proportional to the bunch size for general step-step interactions.
The shape of steadily moving bunches is studied within a continuum theory, 
and analytic expressions for the bunch profile are derived. Scaling laws 
obtained previously for non-moving bunches are recovered asymptotically,
but singularities of the static theory are removed and strong corrections
to scaling are found. The size of the largest
terrace between two bunches is identified as a central
scaling parameter. Our theory applies to a large class of bunching instabilities, 
including sublimation with attachment asymmetry and
surface electromigration in the presence of sublimation or growth.

\end{abstract}

\section{Introduction}

There is much current interest in exploiting 
morphological instabilities to form 
periodic nanoscale patterns at crystal surfaces \cite{Venezuela99,Teichert02}.
Because of their natural in-plane anisotropy,  
vicinal surfaces \cite{Jeong99} prepared at a miscut relative
to a high symmetry orientation provide ideal
substrates for the formation of ripple patterns
parallel or perpendicular to the mean step orientation \cite{Yagi01,Neel03}. 
Here we specifically consider patterns formed by 
\textit{step bunching}, the process in which  
a train of initially equally spaced (straight) steps splits into regions of
densely packed steps (step bunches), and almost flat regions 
\cite{Pimpinelli02, Krug05a}. Bunched semiconductor surfaces are promising templates for 
the growth of metallic nanowires \cite{Roos05}. 

Step bunching can be induced by 
growth \cite{Slanina05,Frisch05}, sublimation \cite{Krug05b}, 
or surface migration of adatoms driven
by an electric current \cite{Yagi01,Yang96,Stoyanov98,Liu98,Sato99,Metois99,Fujita99,Homma00}. 
The common feature of the different instability
mechanisms \cite{Krug05a} is that they break the symmetry between the ascending and 
descending steps bordering a terrace. The appearance of step bunches
thus provides information about the asymmetry of the attachment/detachment
processes at the steps, as well as about the direction of current-induced
adatom migration. Once formed, the shape of a bunch is 
determined by the balance between the destabilizing forces and the repulsive
step-step interactions that act to regularize the step train. As a result,
the bunch shape displays characteristic scaling laws relating e.g. 
its slope and width to the number of steps in the bunch \cite{Stoyanov98,Liu98}. 
These scaling laws are used in the interpretation of experiments to extract the functional form
of the step interactions as well as material parameters such as the
step interaction strength and the electromigration force \cite{Fujita99,Homma00}.       

The large scale properties of step bunches are captured 
by continuum evolution equations for the surface profile \cite{Pimpinelli02},
which can be derived from the underlying discrete step dynamics in a systematic
manner \cite{Krug05b}. The analysis of static (time-independent)
solutions of these equations leads to scaling laws which are in reasonable
agreement with numerical simulation of the discrete step dynamics \cite{Krug05b}.  
However, in the presence of a non-vanishing sublimation or growth flux, step bunches
are moving objects. Because of the high temperatures involved, sublimation -- and hence,
bunch motion -- is significant also in electromigration experiments, where it is 
not the primary cause of bunching \cite{Yang96,Sato99}. 

In this Letter we show that bunch motion alters
the shape and scaling properties of bunches in a fundamental way. 
It removes the artificial symmetry between the in-flow and out-flow regions
(in which steps move into and out of the bunch, respectively) and the concomitant
singularities of the static solutions at the bunch edges \cite{Krug05b}.
We show that the lateral speed of a bunch is inversely proportional to its height
for a large class of models, and we identify the size of the largest terrace
$l_{\mathrm{max}}$ as a natural scaling parameter,  
in terms of which other important bunch
characteristics are expressed in a simple way. 
The maximal terrace size $l_{\mathrm{max}}$ 
is uniquely defined, in contrast to the number of steps in the bunch,
which requires a convention to decide which steps belong to it,
and it is directly accessible experimentally 
by means of reflection electron microscopy (REM) \cite{Metois99}.

\section{Discrete model}

We consider a system of non-transparent steps \cite{Krug05a} 
described on the discrete level by the equations of motion
\begin{equation}
\frac{d x_{i}}{d t} 
=\frac{1-b}{2}\left(  x_{i+1}%
-x_{i}\right)  +\frac{1+b}{2}\left(  x_{i}-x_{i-1}\right)
+U\left(  2f_{i}-f_{i-1}-f_{i+1}\right)  
\label{discrete_time_evolution}
\end{equation}
for the step positions $x_i(t)$, where the time scale has been normalized to the
growth or sublimation flux. The parameter $b$ governs the asymmetry between
ascending and descending steps, relative to the mean step velocity.
The linear form of the first two terms on the right hand side of 
(\ref{discrete_time_evolution}) is characteristic of slow 
attachment/detachment kinetics, and applies equally to step bunching induced by 
sublimation, growth or surface electromigration \cite{Krug05b,Liu98};
here we will assume a sublimating step train going uphill
in the $+x$ direction.  
The last term on the right hand side of (\ref{discrete_time_evolution}) 
represents stabilizing
step-step interactions of strength $U$. In the usual case of entropic or 
dipolar elastic interactions 
\begin{equation}
f_{i}=\left(  \frac{l}{x_{i}-x_{i-1}}\right)  ^{\nu+1}-\left(  \frac{l}%
{x_{i+1}-x_{i}}\right)  ^{\nu+1},
\label{step-step interaction}%
\end{equation}
where $\nu=2$ and $l$ is the average terrace length \cite{Jeong99}. 
Explicit expressions for $b$ and $U$ in terms of 
physical parameters are given below in (\ref{parameters}). 

For $b>0$, (\ref{discrete_time_evolution}) leads to an instability of the 
equally spaced step configuration $x_{i}-x_{i-1}=l$ and its segregation into 
step bunches separated by flat regions. The bunches coarsen slowly
in time by coalescence. We are interested in the final regime of
coarsening with a few big bunches left in the system. In this regime, one can
study a periodic array of identical
bunches, each containing $M$ steps, which satisfy
(\ref{discrete_time_evolution}) with $i=1,2,...M$ and 
the helicoidal boundary conditions 
$ x_{M+1}\equiv x_{1}+Ml $.
It is convenient to consider the comoving step coordinates $y_{i}(t)=x_{i}(t)-lt$, 
in which the center of mass of the step configuation does not move.
In this frame, the stationary trajectory of a step is a periodic function with
some (unknown) period $\tau$, $y_{i}(t)=y_{i}(t+\tau)$. 
Stationarity implies that every
step follows the same trajectory, up to a space and time shift, 
according to  $y_{i+s}(t)=y_{i}(t+\tau s/M)+ls,$ with $s=1,2,...M-1$. Inserting this
into (\ref{discrete_time_evolution}) and setting $\Delta(t) = y_{i+1}(t) - y_i(t)$
we obtain an equation for the stationary step trajectory (in the following we omit subscripts)
\begin{equation}
\frac{d y}{d t}   =\frac{1-b}{2}\Delta(t)+\frac{1+b}{2}%
\Delta(t-\frac{\tau}{M})-l
+U\left[  2f(t)-f(t-\frac{\tau}{M})-f(t+\frac{\tau}{M})\right].
\label{stationary_discrete}%
\end{equation}
This is a differential-difference equation for two periodic 
functions $y(t)$ and $f(t)$, which for the time being will be
treated as independent. 
Expanding the functions in Fourier series with frequencies 
$w_{n}=2\pi n/\tau$ and coefficients $Y_n$ and $F_n$, respectively, 
we obtain from (\ref{stationary_discrete}) 
\begin{equation}
w_{n}=\sin\frac{2\pi n}{M}+2U\left(  1-\cos\frac{2\pi n}{M}\right)
\operatorname{Im}[F_{n}/Y_{n}].
\label{omega_n}%
\end{equation}%
Since (\ref{omega_n}) is valid for any $n$, we can choose $n/M\ll1$ 
for large $M$ and expand (\ref{omega_n}) to obtain the expression
$
\tau=M-2\pi U \operatorname{Im} [n F_{n}/Y_{n}]+O(M^{-1}),
$
which in fact determines the dependence of the bunch velocity on $M$:
In the laboratory frame we have $x_i(t + \tau/M) = x_{i+1}(t) + l (\tau/M - 1)$,
which implies that the whole step configuration shifts by 
$l(\tau/M - 1)$ to the right in time $\tau/M$. Hence the lateral bunch
speed $v$ is given by 
\begin{equation}
\label{vlat}
v/l = 1 - M/\tau = \kappa/M+o(M^{-1}),
\end{equation}
where $\kappa=-2\pi U \operatorname{Im}[n F_{n}/Y_{n}]$ 
has to be determined self-consistently for a given form of
step-step interaction. When $f$ has the usual form
(\ref{step-step interaction}) with $\nu=2$, we find numerically that $\kappa$ 
is a constant proportional to the asymmetry, $\kappa \approx 3 b$, provided that the
asymmetry is not too large, $b \leq 0.5$.

In the following we will see that even
though the bunch velocity decreases with increasing 
bunch size, it cannot be neglected.
According to a scaling argument due to Chernov \cite{Stoyanov98,Chernov61,Slanina05},
the scaling $v \sim M^{-1}$ implies that the 
average bunch size should increase with time as $\sqrt{t}$, which is 
consistent with experiments \cite{Homma00,Yang96} and numerous discrete simulation
\cite{Liu98,Sato99}.

\section{Continuum theory}

The continuum evolution equation  
corresponding to the discrete dynamics (\ref{discrete_time_evolution}) 
reads \cite{Krug05b}
\begin{equation}
\frac{\partial h}{\partial t}+\frac{\partial}{\partial x}\left[
-\frac{b h_{0}^{2}}{2m}-\frac{h_{0}^{3}}{6m^{3}}\frac{\partial m}{\partial
x}+\frac{3Ul^{3}}{2m}\frac{\partial^{2}(m^{2})}{\partial x^{2}}\right] + h_0
= 0, \label{PDE}%
\end{equation}
where $h(x,t)$ is the surface profile, $m(x,t)=\partial h/\partial x > 0$
is the slope, and $h_0$ denotes the height of a single step.  
A periodic array of bunches moving at lateral speed $v$ is obtained 
by the travelling wave ansatz \cite{Slanina05}
$h(x,t) = h(\xi)+\Omega t - h_0 t$,
where $\xi = x - vt$ and the function $h(\xi)$ satisfies the
boundary condition $h(\xi+Ml)=h(\xi)+Mh_{0}$. Inserting this into 
(\ref{PDE}) we find that the vertical excess speed $\Omega$  
is related to $v$ by $\Omega = v m_{0}$, where $m_{0}=h_{0}/l$ is the average slope.
Integrating once the ordinary differential equation for $h(\xi)$ then becomes
\begin{equation}
\Omega\left(  \xi+\xi_{0}-h\right)  +\frac{b}{2}\left(  1-\frac{1}{m}\right)
-\frac{m^{\prime}}{6m^{3}}+\frac{3U}{2m}(m^{2})^{\prime\prime}=0. \label{ODE}%
\end{equation}
We denote derivatives by primes, and measure lateral distances in units of $l$ and
heights in units of $h_{0}$. The phase shift $\xi_{0}$ is a constant of integration
satisfying the condition $\int_{0}^{M}m(\xi)\left[  h(\xi)-\xi-\xi_{0}\right]
d\xi=0$, which follows by multiplying (\ref{ODE}) by $m$, integrating 
over the bunch period, and using the boundary conditions. 
For future reference we note
that, for small $b$, $\Omega \approx 3 b/M$ because of 
(\ref{vlat}). 

\begin{figure}
\twofigures[scale=0.55]{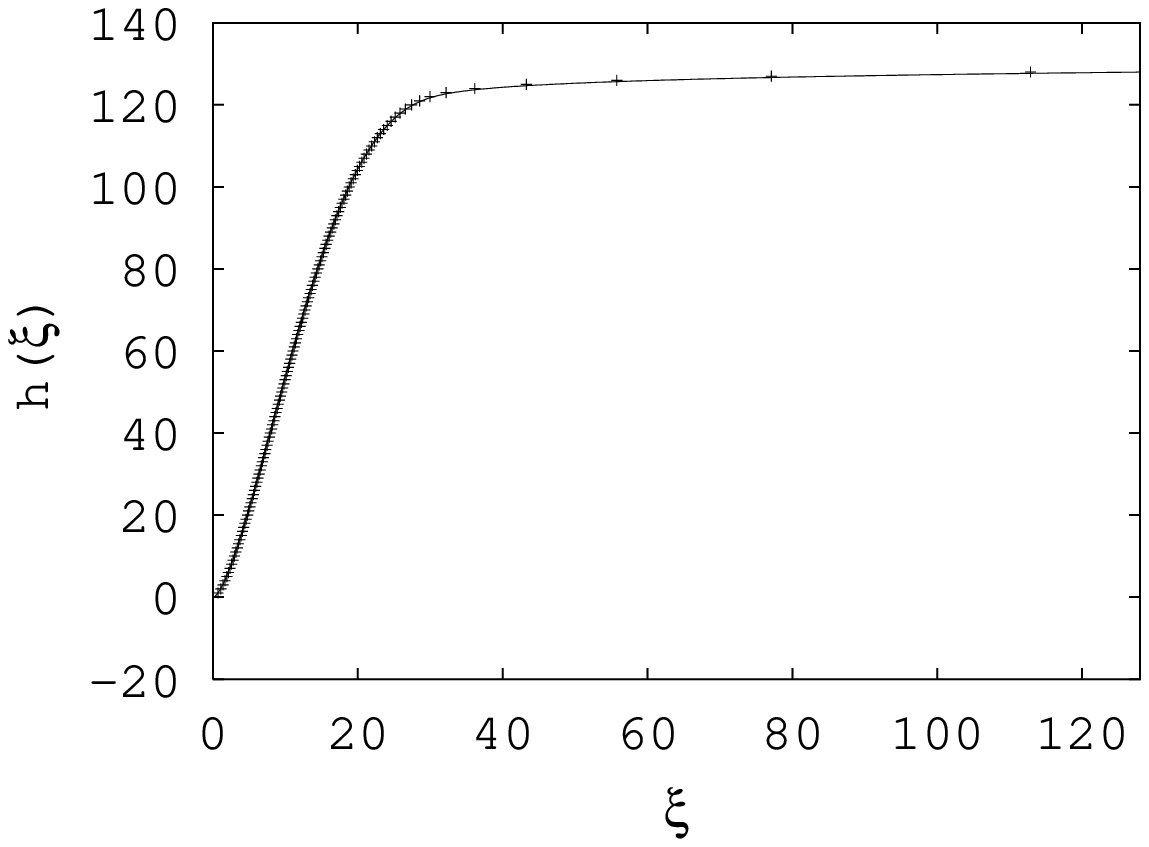}{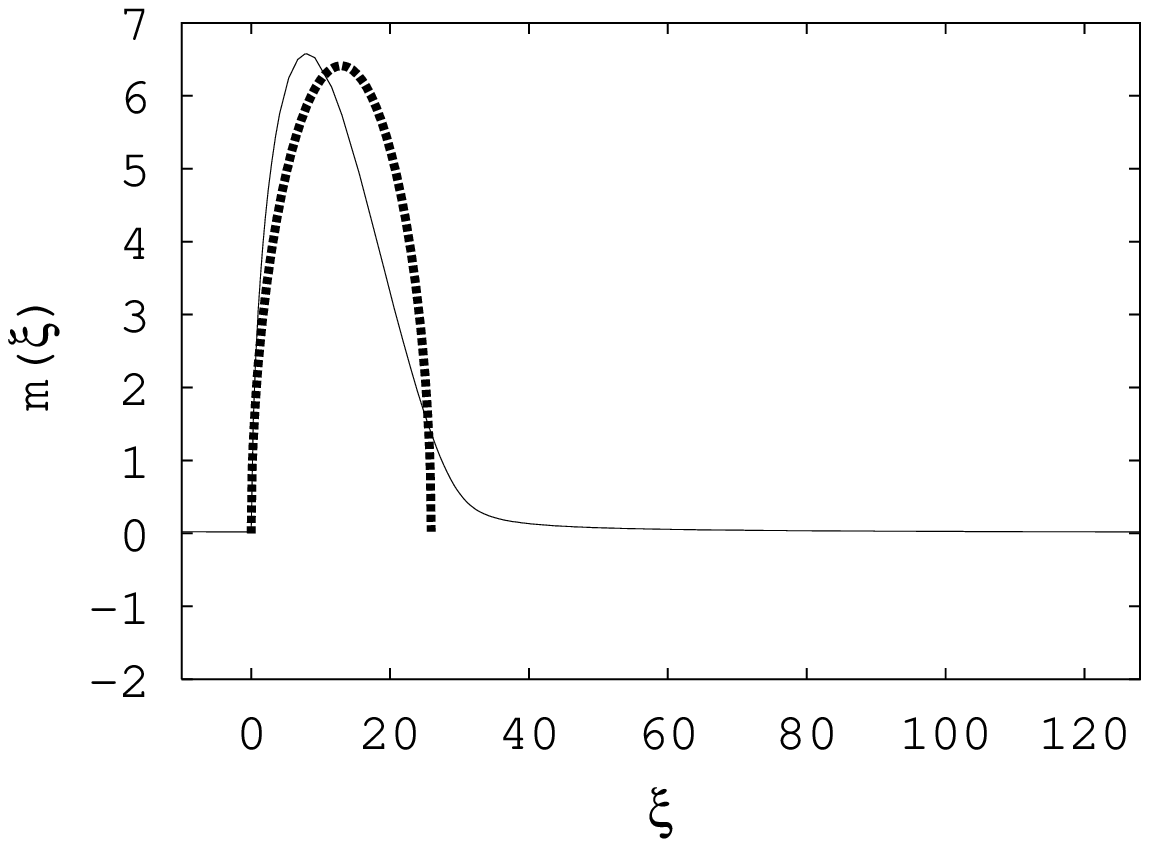}
\caption{Shape of a moving bunch $h(\xi)$ computed from the \textit{discrete} dynamics
(\ref{stationary_discrete}) [symbols],
and the \textit{continuous} evolution equation (\ref{ODE}) 
[full line], for $128$ steps with $b=3/17\approx0.17647,$ and $U=0.569$. There are no
fitting parameters.}
\label{Fig_h(ksi)}
\caption{Slope profile derived from
Fig.\ref{Fig_h(ksi)}, for a moving bunch (full curve) and a static bunch (bold
dashed curve). The slope of the moving bunch increases abruptly in the inflow region and decreases
gradually in the outflow region.}
\label{Fig_Singular_m}
\end{figure}

In Fig.\ref{Fig_h(ksi)} a numerical 
solution of (\ref{ODE}) obtained via a shooting method 
is compared to the discrete step dynamics,
showing excellent agreement. 
The continuum description is generally found to work very well,
provided the asymmetry $b$ is sufficiently small.
In Fig.\ref{Fig_Singular_m} we compare the 
corresponding slope profile to the time-independent solution derived 
in \cite{Krug05b} by setting the terms inside the square brackets in 
(\ref{PDE}) to zero and neglecting the symmetry-breaking term
$(h_0^3/6 m^3) \partial m/\partial x$. As was anticipated in \cite{Krug05b},
the moving bunch is distinctly asymmetric. Moreover the moving solution
extends smoothly over the whole $x$-axis, whereas the static
solution has finite support due to singularities at the bunch
edges. In the following we find analytically the asymptotics of the inflow and outflow regions
of the moving bunch, corresponding to the extreme left and right parts of 
Fig.\ref{Fig_h(ksi)} (see also Fig.\ref{Fig_Singular_m}).

\section{Outflow region: $m(\xi)\ll1$, $m^{\prime}(\xi)<0$} In this region
the steps are far apart and their interaction $U$ is negligible. It is convenient
to perform a Lagrange transform from the function $h(\xi)$ to its inverse $\xi=\xi(h)$. 
After some algebra (\ref{ODE}) with $U = 0$ then reduces to the linear equation
\begin{equation}
\xi^{\prime\prime}-3b\xi^{\prime}+6\Omega\xi=6\Omega(h-\xi_{0})-3b,
\end{equation}
which has the general solution
\begin{equation}
\xi=h-\xi_{0}+C_{1}\exp\left(  \lambda_{1}h\right)  +C_{2}\exp\left(
\lambda_{2}h\right)  \label{complete_outlow_solution}%
\end{equation}
with $\lambda_{1,2}=(3b/2)[1 \pm \sqrt{1-8\Omega/3b^{2}}].$ 
Fixing the boundary conditions so that the point of minimal slope
$\min_\xi m(\xi)=\varepsilon$ is located at $h=0,$ we have two boundary conditions
$\xi(0)=0$ and $\xi^{\prime}(0)=1/\varepsilon$ to determine $C_{1},C_{2}$.
Recalling that $\Omega\approx3b/M$, we see that $\lambda_1 \to 3b$ and 
$\lambda_2 \to 0$ for large bunches, so that (\ref{complete_outlow_solution})
becomes a pure exponential, corresponding to a slope profile
$m(\xi)\approx1/(3b\xi)$. Similar behavior was found in a model with short-range
step interaction, however in that case $\Omega \to 3 b^2/8$ and $\lambda_1 \to 
\lambda_2$ for large bunches \cite{Slanina05}.

\section{Inflow region: $m(\xi)\ll m_{\max}$, $m^{\prime}(\xi)\geq0$} In this
region one can neglect the first two terms in (\ref{ODE}), as can be shown by 
a careful analysis of (\ref{forces_distribution}). The remaining terms give 
$
9U(m^{2})^{\prime\prime}=m^\prime/m^{2}.
$
Integrating once, we find
$
9U(m^{2})^{\prime}=-m^{-1}+\varepsilon^{-1}
$
by requiring the derivative $m^{\prime}(\xi)$ to vanish at the point with
minimal slope $\varepsilon$. Integrating again, we obtain an
implicit equation for $m(\xi)$, 
\begin{equation}
18U\varepsilon\left[  m^{2}/2+m\varepsilon+\varepsilon^{2}\ln(
m/\varepsilon-1) \right]  =\xi.
\label{Singularity_inflow_equation_integrated2}%
\end{equation}
Note that (\ref{Singularity_inflow_equation_integrated2}) is valid for
$m(\xi)>\varepsilon+0$ , to avoid the logarithmic singularity at $m=\varepsilon$. 
In reality there is no singularity, because additional terms 
from (\ref{ODE}) have to be included when $m \to \varepsilon$, which however are completely
irrelevant in the remaining part of the inflow region. 

\section{Scaling laws} We are now prepared to 
investigate the scaling properties of large step bunches ($M\gg1$).

(A) The easiest is to find the \textit{size of the first terrace in the bunch}
$l_{1}$, defined as in \cite{Krug05b} by $m(h=h_{0})=h_{0}/l_{1}$, since
this region is well described by (\ref{Singularity_inflow_equation_integrated2}).
Away from the singularity, for $m \gg \varepsilon$, 
(\ref{Singularity_inflow_equation_integrated2}) reduces to $m(\xi) \approx \sqrt{\xi/9 U 
\varepsilon}$, which is of a similar form as the Pokrovsky-Talapov singularity
found for static bunches \cite{Krug05b}. This yields immediately
\begin{equation}
l_{1}\approx (6U\varepsilon)^{1/3}. \label{first_terrace_size}%
\end{equation}

(B) To estimate the \textit{size of the minimal terrace} in the bunch
$l_{\min}=1/m_{\max}$, we multiply (\ref{ODE}) by $m(m^{2})^{\prime}%
=2m^{2}m^{\prime} = (2/3)(m^3)^{\prime}$ and integrate from $\xi_1$ to $\xi_2$:
\begin{equation}
\left[  \frac{3U}{4}\left(  (m^{2})^{\prime}\right)  ^{2}+b\left(  \frac
{m^{3}}{3}-\frac{m^{2}}{2}\right)  \right]  _{\xi_1}^{\xi_2}=
-2\Omega%
{\displaystyle\int\limits_{\xi_1}^{\xi_2}}
m^{2}m^{\prime}\left(  \xi+\xi_{0}-h\right)  d\xi+\frac{1}{3}%
{\displaystyle\int\limits_{\xi_1}^{\xi_2}}
\frac{(m^{\prime})^{2}}{m}d\xi.
\label{forces_distribution}%
\end{equation}
Setting $\xi_1=0$, $m(0)=\varepsilon$ and $\xi_2=\xi_{\mathrm{max}}$
with $\ m(\xi_{\mathrm{max}})=m_{\max}$, the
left hand side gives $\approx bm_{\max}^{3}/3$ for $m_{\max} \gg 1$. 
The second integral on the
right hand side can be taken, noting that the function $(m^{\prime})^{2}/m$
vanishes everywhere except in the narrow inflow region, where
(\ref{Singularity_inflow_equation_integrated2}) holds, yielding $\int_{0}^{\xi_{\mathrm{max}}}(
(m^{\prime})^{2}/m)  d\xi= \left[  -m^{\prime}-(  36Um^{2})
^{-1}\right]_{0}^{\xi_{\max}}\approx (36U\varepsilon^{2})
^{-1} \equiv I_{0}$. 
To estimate the first integral, first set $\xi_2=M$ in (\ref{forces_distribution})
so that the left hand side vanishes due to the periodic boundary conditions. 
We obtain then $I_{\Omega}\left[  0,M\right]  =2\Omega\int_{0}^{M}m^{2}m^{\prime}\left(
\xi+\xi_{0}-h\right)  d\xi\approx I_{0}/3$. Denoting by $\gamma=\lim
_{M\rightarrow\infty}I_{\Omega}\left[  0,\xi_{\max} \right]  /I_{\Omega}\left[
0,M\right]  $ the relative contribution to the integral from the segment
$\left[  0,\xi_{\max}\right]  $ for large $M$, we find from
(\ref{forces_distribution}) $bm_{\max}^{3}\approx\left(  1-\gamma\right)
I_{0}$, or
\begin{equation}
m_{\max}^{-1}=l_{\min}\approx\left(  
\frac{36U\varepsilon^{2}b}{1-\gamma}\right)^{1/3}. \label{minimal_terrace_size}%
\end{equation}
Numerically we observe that the value of $\gamma$ indeed saturates to a
fixed value for large $M$, and depends rather weakly on $b$ and $U$. Varying
$b$ and $U$ around physically relevant choices of parameters, e.g.,
those in Fig.\ref{Fig_h(ksi)} or in \cite{Krug05b}, we find
$\gamma\approx0.7\pm0.01$. 

(C) \textit{Bunch width }$W$. The definition of the bunch width depends on the
convention used to assign steps to the bunch \cite{Krug05b}. 
Here we define the bunch as the collection of terraces with sizes
smaller than the mean terrace size $l$. 
We can obtain an estimate of $W$ integrating the first term on the 
right hand side of (\ref{forces_distribution})  
by parts: $\int_{0}^{M}m^{2}m^{\prime}\left(  \xi+\xi_{0}-h\right)
d\xi=(1/3)\int_{0}^{M}m^{3}\left(  m-1\right)  d\xi$, and arguing that the
dominant contribution to the integral comes from the bunch interior. Then
$\int_{0}^{M}m^{3}\left(  m-1\right)  d\xi\approx\int_{0}%
^{W}m^{4}d\xi=Qm_{\max}^{4}W\approx MI_{0}/(6b)$, where $Q<1$, and we have used
$\Omega\approx3b/M$. Substituting $m_{\max}$ we get
\begin{equation}
W=\frac{Q^{-1}M}{6\left(  1-\gamma\right)  ^{4/3}}\left(  36U\varepsilon
^{2}b\right)  ^{1/3}. \label{W}%
\end{equation}
For the parameters of Fig.\ref{Fig_h(ksi)}, $Q\approx0.3237$ for large $M$, and
it changes only slightly ($\pm2\%$) under significant variations of $b$ and
$U$.

(D) \textit{The minimal slope.} As we have seen, many     
characteristics of the moving bunch are controlled by the 
single parameter $\varepsilon$,
which can be defined microscopically as the inverse size
of the \textit{largest terrace in the outflow region} between two consecutive
bunches, $l_{\max}=1/\varepsilon$. In order to make a connection to earlier
studies, we need the dependence of $\varepsilon$ on the bunch size $M$.
From the asymptotic slope profile $m \approx 1/(3 b \xi)$ in the outflow
region, one expects $\varepsilon \approx (3bM)^{-1}$ to
leading order. Numerical studies suggest however strong finite size
corrections even for large bunches, $M\lesssim200$. The behaviour
of $\varepsilon$ is rather well approximated by $\varepsilon^{-1}%
\approx3b\alpha M-A(b,U)$ where \ $\alpha\lesssim1$ and typically $A$ is not
small, e.g., for $b,U$ from Fig.\ref{Fig_h(ksi)}, $A$ $\approx13$ in the range
of bunch sizes $50\leq M\leq550$. 

If we nevertheless use the asymptotic expression $\varepsilon \approx 
(3 b M)^{-1}$ in (\ref{first_terrace_size})-(\ref{W}) we
recover the scaling laws 
derived in \cite{Krug05b} for static bunches, however with different numerical prefactors. 
This provides an \textit{a posteriori} justification for the 
agreement between the predictions for static bunches and the numerical
data in \cite{Krug05b}.  
Noting that the dimensionless parameter $S$ introduced in 
\cite{Krug05b} is given by $S = U/(bl)$ in the present units, we see
that (\ref{first_terrace_size}) and (\ref{minimal_terrace_size}) 
reduce to $l_1 \approx (2 S/M)^{1/3}$ and $l_{\min} \approx        
(13.3 \times S/M^2)^{1/3}$, which is to be compared with the 
expressions $l_1 = (4 S/M)^{1/3}$ and $l_{\min} = (16 S/M^2)^{1/3}$
for static bunches. From (\ref{W}) we find that 
the ratio $W/(M l_{\min}) = [6 Q (1 - \gamma)]^{-1} \approx 1.72$, 
which is considerably larger than the corresponding number 1.29
in the static case. This reflects the fact that moving bunches are
considerably broader than their static counterparts, because of the
gradual increase of the terrace sizes in the outflow region, and explains
the significant discrepancy between numerical and analytic estimates
for $W$ in \cite{Krug05b}. 
Finally, we note that for a general
step-step interaction exponent $\nu$ in (\ref{step-step interaction})
we arrive at generalized scaling laws $l_{1}\sim\varepsilon^{1/\left(
\nu+1\right)  },l_{\min}\sim\varepsilon^{2/\left(  \nu+1\right)  }$ and
$W\sim\varepsilon^{-1/\left(  \nu+1\right)  }$ which are, apart from strong
finite size corrections, consistent with the ones derived in
\cite{Krug05b}.

\section{Experimental considerations}

An important condition for the applicability
of our continuum theory is the smallness of the asymmetry parameter, 
$b \leq 0.5$. 
For step bunching induced by a conventional Ehrlich-Schwoebel (ES) effect
during sublimation, $b = (k_+ - k_-)/(k_+ + k_-)$, where $k_+$ ($k_-$) is the kinetic
coefficient for attachment to a step from the lower (upper) terrace \cite{Krug05b};
keeping $b$ small then simply requires
a weak ES effect.
For current-induced step bunching in the attachment/detachment limited regime,
one finds \cite{Liu98} 
\begin{equation}
\label{parameters}
b = \frac{k c_{\mathrm{eq}} a^{2} F \tau_{e}}{k_{\mathrm{B}}T} = 
\frac{\Gamma F \tau_e}{2 a^2 k_{\mathrm{B}}T}, \;\;\;\;
U = \frac{\Gamma g \tau_e m_0^3}{2 k_{\mathrm{B}}T}
\end{equation}
where $k = k_+ = k_-$ denotes the attachment rate of adatoms to steps,
$a^2$ the atomic area, $c_{\mathrm{eq}}$ the equilibrium adatom
concentration, $F$ the electromigration force,
$\tau_e$ the monolayer evaporation time, and $g$ the step interaction
strength. The quantity $\Gamma = 2 k c_{\mathrm{eq}} a^4$ is the mobility
of an isolated step. The model (\ref{discrete_time_evolution})
of non-transparent steps is expected to apply in two of the four temperature regimes
\cite{Yagi01} in which step bunching is observed on Si(111), around 900$^{\textrm{o}}$~C and 
around 1250$^{\textrm{o}}$~C \cite{Metois99,Fujita99,Homma00}. The material 
parameters given in \cite{Yang96,Liu98} lead to the estimate
$b \approx 14$ in the low temperature regime and $b \approx 0.3$ in the high temperature
regime. The latter is presumably an upper bound, since a recent estimate \cite{Saul02} of the 
kinetic length \cite{Krug05a} $l_k = D/k$ (where $D$ is the adatom diffusion coefficient) 
at 1200$^{\textrm{o}}$~C indicates that 
the step mobility $\Gamma$ increases less rapidly with increasing temperature than
was assumed in \cite{Yang96}. The step interaction parameter $U$ depends very 
sensitively on the mean miscut $m_0 = h_0/l$. Taking the interaction
strength to be $g \approx 0.1$ eV/\AA$^2$ at 1250$^{\textrm{o}}$~C, the parameters
given in \cite{Yang96,Liu98} yield $U/l \approx 0.2$ for $l = 50$ nm.

\section{Conclusions}

In this paper we have shown that the continuum equation (\ref{PDE}) faithfully represents
the properties of moving step bunches for sufficiently small values of $b$, and that
it can be used to extract accurate analytic expressions for the bunch shape and
its various characteristic length scales. Our study reveals a central role of the 
point of minimal slope (or maximal terrace size) where the outflow region of one bunch
joins the inflow region of the next. An important ingredient is a Fourier analysis of
the discrete equations of step motion, which yields an inverse dependence of the bunch
speed on bunch size under rather general conditions. Further work is needed to 
analytically derive the coefficient $\kappa$ in (\ref{vlat}), and to understand how
to obtain the bunch speed directly from the continuum equation, possibly by exploiting
a recently proposed analogy to front propagation problems \cite{Slanina05}. In addition,
it seems desirable to study the regime of large $b$, and to investigate the consequences of 
bunch motion for bunch coarsening beyond simple scaling arguments \cite{Slanina05, Chernov61}. 

The conditions assumed in this paper should be realizable in 
electromigration experiments on Si(111) at temperatures around 1250$^{\textrm{o}}$~C.
The gross features of the morphology, such as the maximal terrace size $l_\mathrm{max}$
and the bunch width $W$, could be followed in real time using REM, while for the more
delicate measurements of $l_{\mathrm{min}}$ and $l_1$ STM-studies would be preferable.   

\acknowledgments

This work has been supported by DFG within project KR 1123/1-2. We thank H. Omi, O. Pierre-Louis, 
S. Stoyanov and V. Tonchev for useful discussions.

\end{document}